\documentclass{article}
\usepackage[utf8]{inputenc}
\usepackage{graphicx}
\usepackage{geometry}
\usepackage{kz}
\RequirePackage[colorlinks,linkcolor=blue,citecolor=blue,urlcolor=blue]{hyperref}

\title{Comments on ``A Gibbs sampler for a class of random convex polytopes''}
\author{Kentaro Hoffman, Jan Hannig and Kai Zhang\\
University of North Carolina at Chapel Hill}

\begin{document}

\maketitle
\begin{abstract}
    In this comment we discuss relative strengths and weaknesses of simplex and Dirichlet Dempster-Shafer inference as applied to multi-resolution tests of independence. 
\end{abstract}
\section{Introduction}

We would like to congratulate the authors for their contribution to this longstanding open problem in mathematical statistics. Their clever implementation of MCMC to obtain simplex based Dempster-Shafer (DS) samples for parameters of multinomial distribution and its connection to graph theory is extremely thought provoking.
We expect that the contributions seen in this paper will have impacts for years to come. 

In our comment, we would like to contribute our thoughts particularly on the applications of the various DS approaches to nonparametric tests of independence. Nonparametric dependence detection is a classical statistical problem but recently gains great interest from both statisticians and computer scientists due to its applications in machine learning. See \cite{hoeffding1948non,szekely2007,gretton2007kernel, reshef2011detecting,heller2012consistent,chatterjee2019new,shi2020rate}, and references therein.
 
One general approach in nonparametric tests of independence is the multi-resolution approach. See some recent work by \cite{ma2019fisher,zhang2019bet, lee2019testing,gorsky2018multiscale,zhang2021beauty}. Some advantages of this approach include uniform consistency, minimax optimal power, clear interpretability and efficient computation.

The multi-resolution approach reduces the test of independence to the test of discrete uniformity over multinomial distributions. In this paper, we study the performance of such tests, particularly when the sample size is small compared to the resolution. We show potential gains in power and computational scalability from using an alternative  Dirichlet DS method based on unpublished manuscript of \cite{Lawrence2009ANM}.



\section{Simplex and Dirichlet DS for Multinomial Data}
A crucial part to multi-resolution tests of independence is being able to conduct inference for parameters $\mathbf{p} = (p_1, ... p_k)\in \mathcal P=\{\mathbf p:\ 0\leq p_i,\ \sum_{i=1}^k p_i=1\}$ using observation $\mathbf z=(z_1, ... z_k)$  following multinomial$(n, \mathbf p)$ distribution.
In particular, we will compare two different DS approaches to inference parameters $\mathbf p$.

The first, which we shall refer to as {\em simplex DS}, refers to the method used in the discussed paper. This version of the DS is based on inverting a very natural data generating algorithm proposed by \cite{dempster1966new} and described in detail in Section~2 of the discussed paper. In particular, a sample from the multinomial distribution is generated by sampling ancillary variables uniformly at random on a simplex and selecting the multinomial category depending on which section of the simplex the ancillary variable falls into, see Figure~1a in the discussed paper. To generate DS samples one needs to invert this data generating algorithm, i.e., sample new ancillary variables uniformly and find partitions of the simplex, determining the parameters $\mathbf p$, that would reproduces the observed data. The main challenge is that such partitions of the simplex exist only if the ancillary random variables fall into a rather complicated small polytope. The discussed paper proposes an ingenious Gibbs sampler that does just that. Once such ancillary variable  $u^{(t)}$ is sampled, a linear programming problems need to be solved to obtain the particle $\mathcal{F}(u^{(t)})$ of parameters $\mathbf p$ that are needed for inference.

While this Gibbs sampler offers a theoretically interesting representation using shortest path algorithm on graphs, we find that this algorithm has some drawbacks in terms of scalability. For example,  generating 1000 MCMC samples a 7-Dimensional multinomial takes approximately 1 minute. However,
converting the Gibbs sampler's results $u^{(t)}$ into a convex polytope  $\mathcal{F}(u^{(t)})$ takes nearly 20 times as long as running the Gibbs sampler. This is due to the large number of vertexes in the polytopes that result from the Simplex method. The simulation resulted in polygons with an average of 62 and a max over 250 edges. This places a large computational burden on the simplex method. 

\begin{figure}[H]
    \centering
       \begin{tabular}{|c|c|}
       \hline
        Time Running the Gibbs Sampler &  Time converting MCMC to convex polytope \\ \hline
        2.6 seconds  & 53.6 seconds \\
        \hline
        
    \end{tabular}
    \includegraphics[scale = 0.3]{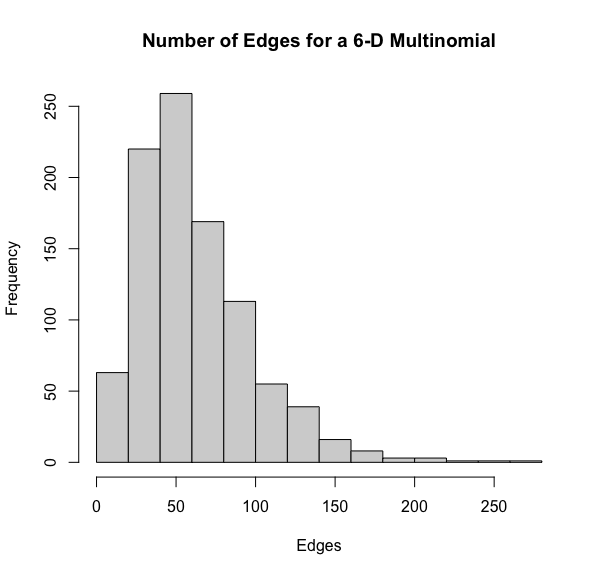}
    \caption{(Top): Runtime comparison of Gibbs sampler vs convex polytope computation. (Bottom): The number of edges in 1000 Convex polytopes from the Simplex method. }
    \label{fig:my_label}
\end{figure}

The other DS method was proposed in an unpublished manuscript of  \cite{Lawrence2009ANM} and we will be referring to as the {\em Dirichlet DS}. While   \cite{Lawrence2009ANM} obtain Dirchlet DS using a clever use of Dempster's rule of combination starting from $k$ binomial DSs. In particular, let $\mathbf W^\star$  follow Dirichlet$(1,z_1,\ldots,z_k)$ distribution and the Dirchlet DS polytope
\begin{equation}\label{eq:DSpolytope}
    \mathcal F(\mathbf W^\star)=\{\mathbf p: W_i^\star\leq p_i,\ i=1,\ldots, k\}
\end{equation}
is a simplex with vertexes $(W_1^\star,W_2^\star,\ldots, W_k^\star)$, 
$(W_1^\star+W_0^\star,W_2^\star,\ldots, W_k^\star)$, $ (W_1^\star,W_2^\star+W_0^\star,\ldots, W_k^\star), \ldots ,$ $(W_1^\star,W_2^\star,\ldots, W_k^\star+W_0^\star)$ \citep{hannig2016}.

Dirichlet DS has a number of clear computational advantages. First, a independent sampling from Dirichlet$(1,z_1,\ldots,z_k)$ can be done directly using methods built into most software packages. Additionally the Dirichlet DS polytope is always a simplex, so for $k=7$ it yields a convex polytope with 8 vertexes, instead of a maximum of 250. This makes the Dirichlet DS to scale to much larger problems, e.g. multinomial distributions with thousands of categories, which is important for testing independence at high dimensions and/or high resolutions.

Next, many estimators of the multinomial proportions have the following invariance property:
If we merge two categories, the estimate of the merged proportion is the sum of the proportions that are being merged. Investigation of \eqref{eq:DSpolytope} reveals that Dirichlet DS has this invariance property. Consequently, inference on proportions for categories in which we have observations is not influenced by addition or deletion of empty categories. As seen in Section~4.1 of the discussed paper, simplex DS does not have this invariance property, but whether this invariance property is desirable could be a matter of opinion.

Finally, let us assume that there were originally $n+r$ of the multinoulli experiments but we were only given results of $n$ and the other $r$ are missing. Notice that we are not assuming that these $r$ multinoullis are missing at random, in fact they could be deleted deliberately. The Dirichlet DS naturally accommodates this additional information by changing the distribution of 
$\mathbf W^\star$ to Dirichlet$(1+r,z_1,\ldots,z_k)$ consequently enlarging the polytope in \eqref{eq:DSpolytope}. Such an enlargement of the DS polytopes is sometimes called weakening. It is not clear to us if simplex DS could be weakened to accommodate missing observations.



\section{DS Test of Independence}
For a pair of continuous random variables $(X,Y)$ distributed on $[0,1]^2$ we would like to test using $n$ i.i.d. samples $(X_1,Y_1)... (X_n, Y_n)$ if $X$ is independent of $Y$. An important step towards this goal is to be able to test if $(X,Y)$ is uniform, or alternatively at what resolutions we have sufficient power to make a rejection. We adopt the notation of \cite{gorsky2018multiscale}. For a joint sample space $\Omega = \Omega_x \times \Omega_y$ it is possible to create a coarse-to-fine discretization of $\Omega$. At resolution $k$, we partition $\Omega$ into sets as follows:
$$\Omega = \bigcup_{i=1}^{k} \bigcup_{j=1}^{k}  I_i ^k \times I_j ^k$$
Where $i \in \{1, k\}$ $I_i^k = [\frac{i-1}{k}, \frac{i}{k})$. At the coarsest level, resolution 2, $\Omega$ is discretized into four different pieces: $\Omega_{11} = [0, \frac{1}{2}] \times [0, \frac{1}{2}]$, $\Omega_{12} = [0, \frac{1}{2}] \times (\frac{1}{2}, 1]$, $\Omega_{21} = (\frac{1}{2}, 1] \times [0, \frac{1}{2}]$, $\Omega_{22} = (\frac{1}{2}, 1] \times (\frac{1}{2}, 1]$ with 
$$\Omega = \Omega_{11} \cup \Omega_{21} \cup \Omega_{12} \cup \Omega_{22}$$
These sets $\{I_i ^k \times I_j ^k \}$ each define statistics $$Z_{i,j} = |(X_i,Y_j) \in \Omega_{i,j}|  $$
which naturally describe a $k \times k $ contingency table. Under the classical multinomial sampling scheme for $k \times k$ contingency table, the statistics $Z_{i,j}$ are distributed:
$$(Z_{11}...Z_{kk} | \sum_{i,j} Z_{i,j}  ) \sim multinomial(n, (p_{11}, ... p_{kk}  )) $$
To test if $(X,Y)$ are uniform at resolution k, it suffices to discretize $(X_1,Y_1) ... (X_n,Y_n)$ into a $k \times k$ contingency table and used the counts $z_{ij}$ in each bin to obtain the distribution of random polytope $\Delta\subset \mathcal P=\{\mathbf p:\ 0\leq p_{ij},\ \sum_{i=1}^k\sum_{j=1}^k p_{ij}=1\}$ following either simplex or Dirichlet DS distribution. The null hypothesis of uniformity should be rejected at this resolution if the null-hypothesis point $(k^{-2},\ldots,k^{-2})$ lies well outside of the bulk of the distribution of $\Delta$. To quantify this we will find the upper and lower probability of the complement of the smallest ball centered on the point estimator $\mathbf{\hat{p}}= \left(\frac{z_{11}+k^{-2}}{n+1},\ldots,\frac{z_{kk}+k^{-2}}{n+1}\right)$ and containing the null-hypothesis point $(k^{-2},\ldots,k^{-2})$. 

To this end we generate $m$ samples $\{ \Delta_1, ...\Delta_m\} $ from either the simplex or Dirichlet DS. Then, we compute the upper and lower distances $\mathcal{U} = \{u_1, ... u_m \}$, $\mathcal{L} = \{l_1, ... l_m \}$, where 
\[l_i = \inf \{\|\mathbf y - \mathbf{\hat{p}}\|_2:\ \mathbf y\in \Delta_i \},\quad
u_i = \sup \{\|\mathbf y - \mathbf{\hat{p}}\|_2:\ \mathbf y\in \Delta_i \}.\]
Leveraging the convexity of $\Delta_i$, $u_i$ can be computed as 
$u_i = \max \{\|\mathbf e - \mathbf{\hat{p}}\|_2:\  \mathbf e \in V(\Delta_i) \},$
where $V(\Delta_i)$ is the set of vertexes of the convex polytope $\Delta_i$, and $l_i$ can be found by solving a convex optimization problem.
We then use $u_i$ and $l_i$ and the distance 
$
r_{center} = \|\mathbf{\hat{p}} - (\frac{1}{k^2}, ... \frac{1}{k^2})\|_2
$
to define upper and lower $p$-values as
\[\hat{p}_{upper}(\{u_1, ... u_m \}) = \frac{|\mathcal{U} \geq r_{center} |}{m},\quad
\hat{p}_{lower}(\{l_1, ... l_m \}) = \frac{|\mathcal{L} \geq r_{center}|}{m}. \]

The upper $p$-value can be then used as usual $p$-value for testing the null hypothesis of uniformity at resolution $k$. The gap between the upper and lower $p$-value is specific to DS and measures lack of knowledge. In particular if the gap between upper and lower $p$-value indicates inability to make decision about independence at this resolution. 

\section{Simulation Results}
To compare the performance of the Simplex and Dirichlet methods, we consider a simple test of independence. First, 100 data sets of sample size $n = 30$ are generated under either the following null or alternate hypotheses:
\begin{equation}\label{eq:Hypotheses}
H_0 \sim  \text{Beta}(1,1)^2,\quad 
H_1 \sim  \text{Beta}(1,2)^2.
\end{equation}
Each of these data sets is then discretized into 2$\times$2,  3$\times$3, and $6\times6$ contingency tables and each table is tested for independence using the simplex DS, the Dirichlet DS, and the classical $\chi^2$ tests. To generate the $p$-values for the test of independence, both the simplex and Dirichlet DS generate 200 polytopes with a burn-in of 300 for the former. The purpose of the low sample size ($n = 30$) in this simulation is to demonstrate that as the resolution $k$ and the number of multinomial categories $k^2$ increases with sample size held  constant, the uncertainty indicated between the gap between the upper and lower $p$-values increases.

\begin{figure}[H]
    \centering
    \includegraphics[scale = 0.5]{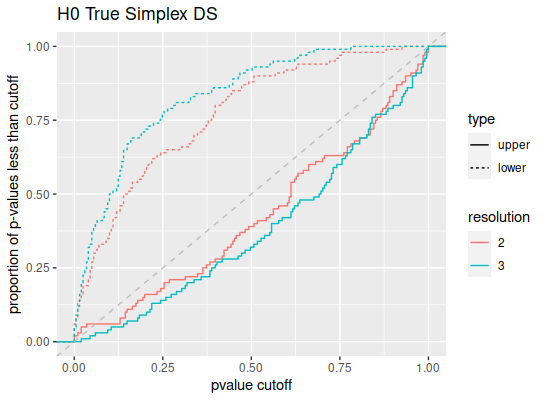}
    \includegraphics[scale = 0.5]{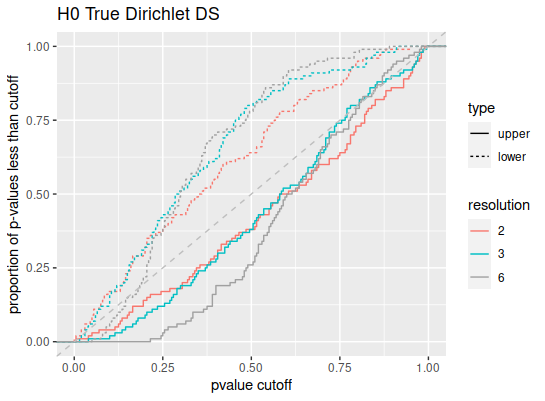}
    \includegraphics[scale = 0.5]{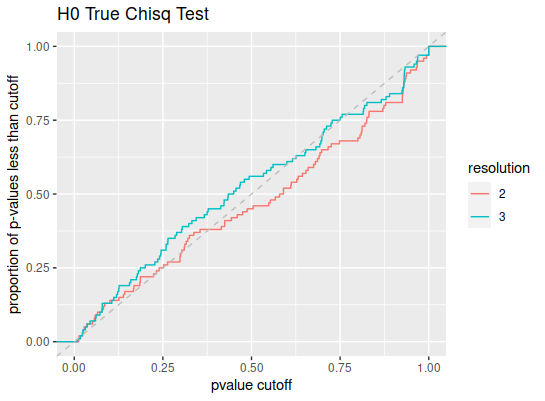}
     \caption{Empirical CDFs of the upper and lower $p$-values for $H_0$ analyzed using three tests: (Top Left):  Simplex DS, (Top Right) Dirichlet DS, and (Bottom): $\chi^2$. The $x$-axis is the nominal $p$-value, the $y$-axis is proportion of $p$-values below $p$-value cutoff.}
    \label{fig:True_Power}
\end{figure}

In Figure \ref{fig:True_Power} we present plots of the Empirical CDFs of the upper and lower $p$-values under the assumption that $H_0$ is true. Well calibrated $p$-values follow a uniform distribution which CDF is represented by the $45^\text{o}$ line. As expected, we see that $p$-values empirical CDFs from the $\chi^2$ test closely follow this dotted line. The upper $p$-values for both DS tests are below the dotted line, showing that these $p$-values are conservative, i.e., sub-uniform. Next, we see that while the upper $p$-values for the Dirichlet and simplex method behave similarly, the lower $p$-values of the simplex method are more skewed towards rejecting. Consequently, Dirichlet DS has a much smaller gap between the upper and lower $p$-values than the simplex DS. Finally, we remark that there are no $p$-values for the 6x6 simplex method as the computation timed out after 2hr without producing a simplex.

\begin{figure}[H]
    \centering
    \includegraphics[scale = 0.5]{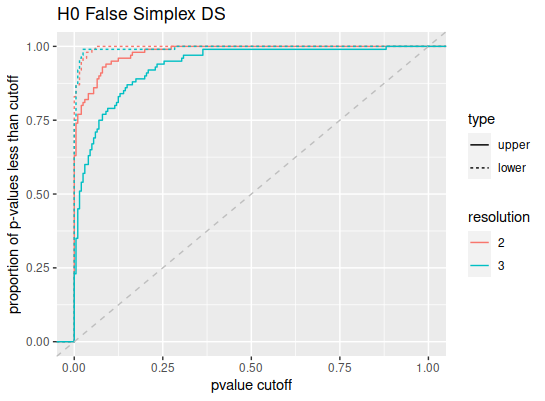}
    \includegraphics[scale = 0.5]{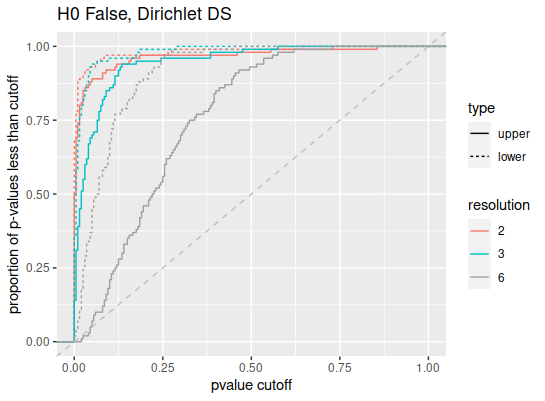}
    \includegraphics[scale = 0.5]{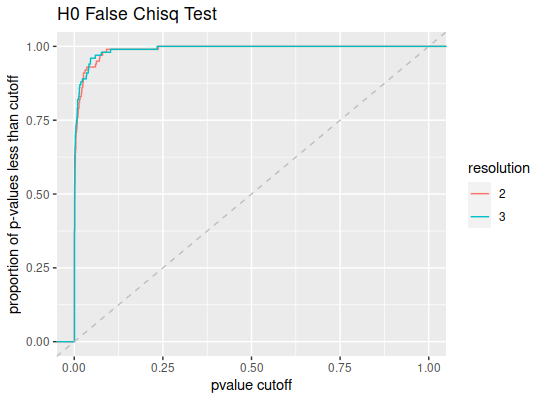}
    \caption{
    Empirical CDFs of the upper and lower $p$-values for $H_0$ analyzed using three tests: (Top Left):  Simplex DS, (Top Right) Dirichlet DS, and (Bottom): $\chi^2$. The $x$-axis is the nominal $p$-value, the $y$-axis is proportion of $p$-values below the cutoff.}
    \label{fig:False_power}
\end{figure}

In Figure \ref{fig:False_power} we show empirical CDFs based on data generated under the alternate hypothesis in \eqref{eq:Hypotheses}. All three tests correctly lean towards rejecting the null hypothesis. In terms of the power of the lower $p$-value, the simplex method performs similarly to the Dirichlet method. However, the empirical CDFs of upper $p$-values for the simplex method is lower than their corresponding empirical CDFs for the Dirichlet DS empirical Cr. This indicates that for the Dirichlet method has more power to reject $H_0$. In addition, we can see the gaps between lower and upper $p$-value plots increase as the resolution increases in both the simplex and Dirichlet DS. 
\\\\
As for runtime comparisons, the difference is substantial. Generating one polytope under Dirichlet DS at the $3 \times 3$ level takes approximately 2 seconds while a similar polytope takes nearly 30 seconds under Simplex DS. The difference in runtime comparison gets larger with the $6\times6$ level, where the Dirichlet DS is still under 5 seconds while the Simplex DS timed out after at least 2 hours.
 \begin{table}
     \centering
     \begin{tabular}{|c|c|}
\hline
     \textbf{Method} & \textbf{Runtime to generate one polytope}  \\ \hline 
     Dirichlet Method for $2\times2$ & 1.76 seconds  \\ \hline
     Dirichlet Method for $3\times3$ & 2.15 seconds  \\ \hline
     Dirichlet Method for $6\times6$ & 4.15  seconds \\ \hline
     Simplex Method $2\times2$ &  5.31 seconds \\ \hline
     Simplex Method $3\times3$ &  29.60 seconds \\ \hline
     Simplex Method $6\times6$ &  Timed out at $>2$hr \\ \hline
\end{tabular}
     \label{tbl:my_label}
 \end{table}

\section{Discussion}
Based on this small simulation example, it appears that the Dirichlet DS might be better suited for performing the test for independence than the simplex DS both in terms of speed and frequentist performance. Dirichlet DS also has a relatively easy way to adapt to  situations with missing data with a potential to be applied to adversarial attack scenarios. The one negative is that unlike simplex DS, the Dirichlet DS does not appear to be fiducial, i.e., based on an inverse of a data generating algorithm. On balance, it is not clear to us in what practical situations the simplex DS would be the preferable choice over the Dirichlet DS when conducting statistical inference.

\section*{Acknowledgements}
Jan Hannig’s research was supported in part by the National Science Foundation under Grant No. IIS-1633074 and DMS-1916115. Kai Zhang's research was supported in part by the National Science Foundation under Grant No. DMS-1613112, IIS-1633212, and DMS-1916237.

\bibliographystyle{Chicago}
\bibliography{KZ_BIB_210228}

\end{document}